\documentclass[twocolumn,showpacs,showkeys,amsmath,amssymb]{revtex4}
\usepackage{graphicx}
\usepackage{epsfig}
\usepackage{dcolumn}
\usepackage{bm}

\begin{document}

\title{Discovery of New Physics in Radiative Pion Decays?}

\author{M. V. Chizhov}
\affiliation{
Centre of Space Research and Technologies, 
University of Sofia, 1164 Sofia, Bulgaria}


\begin{abstract}
Recently a strong indication for a deviation from the standard model (SM)
has been obtained by PIBETA Collaboration. Namely, SM fails to describe 
the energy distribution and the branching ratio of the radiative decays 
of the positive pions at rest in the high-$E_\gamma$/low-$E_e$ kinematic region. 

The previous experiment at ISTRA facility, testing the radiative decays 
of negative pions in flight in a wide kinematic region has alarmed about 
the same effect, although statistically less significant. 
The present PIBETA result indicates a deficit of the branching ratio 
of the radiative pion decay in the specified kinematic region at $8\sigma$ 
level in comparison with SM prediction, while in the other kinematic regions 
both the branching ratios and the energy distributions are compatible with 
the \mbox{$V\!-\!A$} interaction.  

We argue that this effect can result only from a small admixture of new tensor 
interactions. They may arise due to an exchange of new spin one {\em chiral bosons} 
which interact {\em anomalously} with matter.
\end{abstract}


\maketitle

\section{Introduction}

The standard model (SM) includes three different types of 
fundamental interactions: electromagnetic, weak and strong ones.
The corresponding forces arise due to an exchange of spin-1 
gauge bosons, described by four-vector fields $V_\alpha$.
For the completeness of SM at least one more scalar boson is required.
However, due to its weak coupling to the ordinary matter
and its big mass, this particle has not been detected yet
by experiments.

Meanwhile, the existence of many other bosons has been predicted
theoretically and tested experimentally. The supersymmetry, for example,
suggests  a rich variety of new particles, which, however, 
have not been yet observed  experimentally. 

In this letter we show that the recent PIBETA result~\cite{PIBETA} 
for the radiative pion decays (RPD) $\pi^+\to e^+\nu\gamma$
indicates that a different kind of fundamental spin-1 {\em chiral bosons}
may be present in Nature. They have not been discussed 
intensively in literature till now. 

These particles were first mentioned as a different type of spin-1 bosons 
in ref.~\cite{Kemmer}. They appear naturally in the analysis of possible 
Yukawa couplings of~spin-1 bosons to fermion currents. In the relativistic 
physics, where two different types of spin-1/2 fermions exist, namely left-handed 
$\psi_L=\frac{1}{2}(1-\gamma^5)\psi$ and 
right-handed \mbox{$\psi_R=\frac{1}{2}(1+\gamma^5)\psi$},
two different types of Lorentz invariant interactions are possible: 
{\it \`a la gauge} vector interactions
\begin{equation}
{\cal L}_V=\left(g_{LL}~\overline{\psi_L}\gamma^\alpha\psi_L+
g_{RR}~\overline{\psi_R}\gamma^\alpha\psi_R\right)V_\alpha
\label{vector}
\end{equation}
 and tensor interactions 
\begin{equation}
{\cal L}_T=g_{LR}~\overline{\psi_L}\sigma^{\alpha\beta}\psi_R T^+_{\alpha\beta}+
g_{RL}~\overline{\psi_R}\sigma^{\alpha\beta}\psi_L T^-_{\alpha\beta},
\label{tensor}
\end{equation}
where $\sigma^{\alpha\beta}=\frac{i}{2}[\gamma^\alpha,\gamma^\beta]$,
$T^\pm_{\alpha\beta}$ are antisymmetric tensor fields and $g$ denotes 
dimensionless coupling constants. 

Both the vector fields $V_\alpha$ and the tensor fields $T^\pm_{\alpha\beta}$ 
describe particles with spin one. However, they interact absolutely 
differently with matter. The gauge particles are chirally neutral and, hence, they
preserve the helicities of incoming and outgoing fermions. On the contrary,
the tensor bosons carry a {\em chiral charge} that leads to helicity flip for
the interacting fermions. Moreover, self-interactions of the {\em chiral bosons} exist  
even for an abelian case, leading to a negative contribution
into the~$\beta$-function~\cite{Avdeev}.  

An example of the presence of the new kind of {\em chiral particles} 
in Nature is the existence of the axial-vector meson resonance $b_1(1235)$, 
which has only anomalous tensor interactions (\ref{tensor}) with quarks. 
The successful description of the dynamical properties of the hadron systems~\cite{NJL} 
hints that the same phenomenon may have place at the electroweak
scale as well, similarly to the gauge-like hadron resonances $\rho$ and $a_1$,
which have served as prototypes for the photon and the weak bosons~\cite{Weinberg}.

The PIBETA result confirms the anomaly already observed at the ISTRA facility
\cite{ISTRA} for the negative pion decays $\pi^-\to e^-\tilde{\nu}\gamma$ in flight. 
It is remarkable, that these two experiments with absolutely different systematics 
revealed the same effect: the deficit in PRD yield. Moreover, the PIBETA experiment
has measured simultaneously absolute total $\pi\to e \nu$, $\pi^+\to\pi^0 e^+\nu$,
$\mu\to e\nu\tilde{\nu}$, as well as partial $\mu\to e\nu\tilde{\nu}\gamma$ branching 
ratios, which are in excellent agreement with SM predictions, with better than 1\% 
accuracy~\cite{Emil}. The only deviation from SM is observed in the 
high-$E_\gamma$/low-$E_e$ kinematic region exactly where the effect has been 
predicted~\cite{Poblaguev,MPL}. The effect is so big that it cannot be explained 
within SM~\cite{Kuraev} and its supersymmetric extensions~\cite{SUSY}. 

Historically various types of {\em local} four-fermion effective interactions 
have been introduced to test the eventual presence of a new physics. 
Here we will follow also the phenomenological approach of the effective interactions
because the complete theory for the interacting tensor particles is not 
constructed yet. However, we argue that the effective interactions should include
new {\em non-local} momentum dependent tensor interactions, in order to 
describe the PIBETA result without contradiction with the present experimental data.

\section{New tensor interactions}

First of all we define the Lorentz structure 
of the new interactions  on the basis of PIBETA and ISTRA results.
Since both collaborations have observed a deficit  
of events in comparison with SM expectation, it should stem from 
destructive interference between SM and the new interactions in RPD.

It is well known~\cite{Bryman} that the dominant contribution 
in RPD comes from the inner bremsstrahlung (IB) process 
near the edge $E_\gamma+E_e\simeq m_\pi/2$ of the kinematically allowed region,
which include the high-$E_\gamma$/low-$E_e$
kinematic region, where the deficit of the events was observed.
We assume that the new interactions interfere destructively
with this process leading to the deficit of events.
Hence, the new interactions should have the same chiral structure
as IB which is different from \mbox{$V\!-\!A$} interactions. Therefore, 
they do not interfere with the latter due to the smallness of the electron mass. 
This is one of the reasons why such type of interactions are hard to  
detect and why they have not been observed before. 

Let us consider all possible current-current interactions
which lead to a helicity flip.
There are only three possible Lorentz structures obeying this property: 
scalar, pseudoscalar and tensor currents. The matrix element of the
pion-photon transition for the scalar quark current
$\langle\gamma\vert\bar{u}d\vert\pi\rangle$ is zero by kinematic
reasons and does not contribute to RPD. On the other hand the contribution
of the pseudoscalar quark current to the matrix element of the ordinary
pion decay
\begin{equation}
i\langle 0\vert\bar{u}\gamma^5d\vert\pi\rangle=
\frac{m^2_\pi}{(m_d+m_u)m_e}m_e f_\pi\simeq 3.8\times 10^3 ~m_e f_\pi 
\label{pseudoscalar}
\end{equation}
is enormously enhanced in comparison with the standard chirally suppressed 
\mbox{$V\!-\!A$} contribution, and it is severely constrained 
by the experimental data~\cite{R}. On the contrary, the matrix element of
the tensor quark current
$\langle 0\vert\bar{u}\sigma_{\alpha\beta}d\vert\pi\rangle$ is zero
by kinematic reasons, 
and does not contribute directly to the pion decay, 
thus escaping the experimental constraints.

Hence, only less constrained tensor interactions 
are possible candidates for the  explanation of RPD anomaly.
We argue that particularly non-local momentum dependent tensor interactions
are responsible for the detected anomaly. 

In order to explain the ISTRA anomaly a new quark-lepton tensor
interactions
\begin{equation}
{\cal L}^{loc}_T = -\sqrt{2} f_T G~ 
\bar{u}\sigma_{\alpha\beta}d~
\bar{e}\sigma_{\alpha\beta}\nu_L+{\rm h.c.},
\label{LT1}
\end{equation}
with the effective coupling constant $f_T\simeq 10^{-2}$ have been 
introduced~\cite{Poblaguev}, here $G=G_F V_{ud}$.
The dimensionless coupling constant $f_T$ determines the strength of 
the new tensor interactions relative to the ordinary weak interactions,
governed by the Fermi coupling constant $G_F$. 
Although such tensor interactions, introduced ad hoc, can explain 
the \mbox{ISTRA} anomaly, it has been pointed out~\cite{Voloshin}
that the necessary value of the coupling constant $f_T$
contradicts the ordinary pion decay $\pi\to e\nu$. 
This happens because owing to the electromagnetic radiative corrections 
the pseudotensor quark current $\bar{u}\sigma_{\alpha\beta}\gamma^5 d$
leads to a generation of the pseudoscalar quark
current $\bar{u}\gamma^5 d$, to which pion decay is very sensitive.

The generation of the pseudotensor term $\bar{u}\sigma_{\alpha\beta}\gamma^5 d$ 
cannot be avoided for derivative free local four-fermion interactions even if we have 
started with a parity conserving quark current $\bar{u}\sigma_{\alpha\beta}d$ 
in eq.~(\ref{LT1}).
Owing to the identity
$\bar{u}\sigma_{\alpha\beta} d_R~\bar{e}\sigma_{\alpha\beta}\nu_L\equiv 0$,
the Lagrangian (\ref{LT1}) effectively reads
\begin{equation}
{\cal L}^{loc}_T = -\sqrt{2} f_T G~ 
\bar{u}\sigma_{\alpha\beta} d_L~
\bar{e}\sigma_{\alpha\beta}\nu_L+{\rm h.c.},
\end{equation}
where the chiral structure shows itself in the quark current.

The solution of this problem was found in ref.~\cite{MPL} via introducing non-local
momentum dependent tensor interactions
\begin{eqnarray}
{\cal L'}_T=&-&\sqrt{2}f_T G~\bar{u}\sigma_{\lambda\beta}d_L~
\bar{e}\sigma_{\lambda\beta}\nu_L
\nonumber\\
&-&\sqrt{2}f'_T G~\bar{u}\sigma_{\lambda\alpha}d_R~
\frac{4Q_\alpha Q_\beta}{Q^2}~
\bar{e}\sigma_{\lambda\beta}\nu_L+{\rm h.c.} 
\nonumber\\
=&-&\sqrt{2}f'_T G~ 
\bar{u}\sigma_{\lambda\alpha}d~
\frac{4Q_\alpha Q_\beta}{Q^2}~
\bar{e}\sigma_{\lambda\beta}\nu_L+{\rm h.c.},
\label{R}
\end{eqnarray}
where $Q_\alpha$ is the momentum transfer between quark and lepton currents,
and $f'_T=f_T$ are positive dimensionless coupling constants.
In this case the second term in the second row of eq.~(\ref{R}) 
is no longer equal to zero, despite the different chiral structures 
in quark and lepton currents.
And due to the identity
\begin{equation}
\bar{u}\sigma_{\alpha\beta}d_L~\bar{e}\sigma_{\alpha\beta}\nu_L\equiv 
\bar{u}\sigma_{\lambda\alpha}d_L~
\frac{4Q_\alpha Q_\beta}{Q^2}~
\bar{e}\sigma_{\lambda\beta}\nu_L,
\label{qLR}
\end{equation}
it can compensate the opposite chiral quark structure of the first term. 
Then the terms with the pseudotensor quark currents 
$\bar{u}\sigma_{\alpha\beta}\gamma^5 d$ cancel out in eq.~(\ref{R}), 
and the tensor current 
$\bar{u}\sigma_{\alpha\beta} d$ does not contribute to pseudoscalar pion
decay because of parity conservation in electromagnetic interactions.

The two different terms in the effective Lagrangian (\ref{R}) come from exchanges
of new spin-1 bosons $T^\pm_{\alpha\beta}$ and $U^\pm_{\alpha\beta}$ 
with opposite chiral charges, which are necessary
to avoid a chiral anomaly~\cite{MPL}.
The peculiar Lorentz structure of the new tensor interactions reflects
the Lorentz structure of the propagators for the chiral bosons.
This structure can be obtained, for example, from the one-loop radiative 
corrections to the self-energy of the tensor fields using the interactions 
(\ref{tensor}). This follows from the fact that in case of dimensionless 
coupling constants the theory is formally renormalizable 
and the radiative corrections should reproduce 
the Lorentz structure of the kinetic terms~\cite{Landau} 
in the bare initial Lagrangian for the tensor fields.

In general, the coupling constants $f_T$ and $f'_T$ can be different,
however, we assume their equality
to avoid the experimental constraint from the ordinary pion decay.
In the following we will keep different notations for the coupling
constants in order to compare the effects from the two different Lagrangians
(\ref{LT1}) and (\ref{R}).

\section{Radiative pion decay}

The most general matrix element of RPD $\pi^-\to e^- \tilde{\nu}\gamma$
reads
\begin{equation}
M=M_{IB}+M_{SD}+M_T+M'_T,
\label{M}
\end{equation}
where besides SM matrix elements for IB process
\begin{equation}
M_{IB}=i\frac{eG}{\sqrt{2}}f_\pi m_e\varepsilon_\alpha~
\bar{e}\left[\frac{2p_\alpha}{pq}
-\frac{2k_\alpha-i\sigma_{\alpha\beta}q_\beta}{kq}\right]\nu_L
\label{IB}
\end{equation}
and for the structure-dependent (SD) radiation 
\begin{eqnarray}
M_{SD}=i\frac{\sqrt{2}eG}{m_\pi}\varepsilon_\alpha \!\!\!&\{&
F_A\left[(pq) g_{\alpha\beta}-p_\alpha q_\beta\right]
\nonumber\\
&+& iF_V\epsilon_{\alpha\beta\rho\sigma}p_\rho q_\sigma\}~
\bar{e}\gamma_\beta\nu_L,
\label{SD}
\end{eqnarray}
the new tensor contributions
\begin{equation}
M_T = -\sqrt{2}eG F_T~ \varepsilon_\alpha q_\beta~
\bar{e}\sigma_{\alpha\beta}\nu_L
\label{MT1}
\end{equation}
and 
\begin{equation}
M'_T = -\sqrt{2}eG F'_T~\left(q_\alpha\varepsilon_\lambda-
q_\lambda\varepsilon_\alpha\right)
\frac{Q_\lambda Q_\beta}{Q^2}~
\bar{e}\sigma_{\alpha\beta}\nu_L,
\label{MT2}
\end{equation}
are present.
Here $\varepsilon_\alpha$ is the photon polarization vector; $p$,~$k$,
and $q$ are pion, electron and photon momenta, correspondingly.

The first term in eq.~(\ref{M}) $M_{IB}$ describes a gauge invariant QED process
of the photon radiation from the external charged particles, which is free from 
the effects of the strong interactions. It contains only one well-known 
phenomenological parameter: the pion decay constant $f_\pi=(130.7\pm 0.4)$ MeV. 

The second term $M_{SD}$ corresponds to the photon emission from hadronic
intermediate states, governed completely by the strong interactions
physics. It is parametrized by the two form factors $F_V$ and $F_A$ of the 
$\pi$-$\gamma$ matrix elements for the vector quark current
\begin{equation}
\langle\gamma(q)|\bar{u}\gamma_\alpha d|\pi^-(p)\rangle =
-\frac{e}{m_\pi}\varepsilon_\beta~
F_V\epsilon_{\alpha\beta\rho\sigma}p_\rho\ q_\sigma
\end{equation}
and for the axial-vector quark current
\begin{eqnarray}
\langle\gamma(q)|\bar{u}\gamma_\alpha\gamma^5 d|\pi^-(p)\rangle\!\!&=&\!\!
i\frac{e}{m_\pi}\varepsilon_\beta
F_A\left[(pq)g_{\alpha\beta}-q_\alpha p_\beta\right]
\nonumber\\
\!\!&+&\!\!
ie\varepsilon_\alpha~f_\pi.
\end{eqnarray}

Assuming CVC hypothesis, the vector form factor $F_V$ is directly related to the
$\pi^0\to\gamma\gamma$ amplitude~\cite{CVC} and can be extracted from
the experimental width of this decay
\begin{equation}
F_V=\frac{1}{\alpha}\sqrt{
\frac{2\Gamma(\pi^0\to\gamma\gamma)}{\pi m_{\pi^0}}}=
0.0262\pm 0.0009.
\end{equation}
This value is in a fair agreement with the calculations in the relativistic quark 
model~\cite{Adler} and with the leading order calculations
of the chiral perturbation theory (CHPT) \cite{p4}
\begin{equation}
F_V=\frac{1}{4\pi^2}\frac{m_\pi}{F_\pi}\approx 0.0270 .
\end{equation}

The value of the axial form factor $F_A$ is model dependent and its
determination is a matter of experimental measurements.
The ratio of the axial to the vector form factors
$\gamma=F_A/F_V$ has been measured in the previous experiments~\cite{old} 
in kinematic regions where the contribution of the new tensor terms is not essential. 
The average value $\gamma=0.448\pm 0.062$ at fixed $F_V=0.0259\pm 0.0005$~\cite{PDG}
is also in agreement with the calculations in CHPT~\cite{CHPT}. 

The matrix elements $M_T$ and $M'_T$ follow from the new interactions between
quark and lepton tensor currents (\ref{R}). The matrix element for the quark
tensor current 
\begin{equation}
\langle\gamma(q)|\bar{u}\sigma_{\alpha\beta}\gamma^5 d|\pi^-(p)\rangle =
-\frac{e}{2} F^0_T\left(q_\alpha\varepsilon_\beta-q_\beta\varepsilon_\alpha\right). 
\end{equation}
can be calculated~\cite{SUSY} 
applying the QCD sum rules techniques and the PCAC hypothesis. So
\begin{equation}
F^0_T=\frac{2}{3}\frac{\chi\langle 0|\bar{q}q|0\rangle}{f_\pi}\approx 0.4
\end{equation}
is expressed through the magnetic susceptibility~\cite{chi,84} 
$\chi=-(5.7\pm 0.6)$~GeV$^{-2}$ of the quark condensate
and its vacuum expectation value
$\langle 0|\bar{q}q|0\rangle\approx -(0.24~{\rm GeV})^3$.
Then the tensor form factors in eqs.~(\ref{MT1}) and (\ref{MT2})
read $F_T=-(f_T+f'_T)F^0_T$ and $F'_T=-2f'_T F^0_T$.

In general, all form factors depend on the square of momentum transfer to
the lepton pair $Q^2=(p-q)^2$. However, these dependences are weak and,
hence, the form factors can be assumed as constants.

The differential decay width of RPD
\begin{equation}
\frac{{\rm d}^2\Gamma_{\pi\to e\nu\gamma}}{{\rm d}x{\rm d}y}=
\frac{\alpha}{2\pi}~
\frac{\Gamma_{\pi\to e\nu}}{(1-r)^2}~\rho(x,y)
\end{equation}
can be expressed through the ordinary pion decay width $\Gamma_{\pi\to e\nu}$, 
where the kinematic variables $x=2pq/m^2_\pi$, $y=2pk/m^2_\pi$ and the ratio 
$r=(m_e/m_\pi)^2\approx 1.34\times 10^{-5}$ are introduced. The Dalitz plot 
distribution is defined by the density 
\begin{eqnarray}
\rho(x,y)&=&\rho_{IB}(x,y)+\rho_{SD}(x,y)+\rho_{IBSD}(x,y)
\nonumber\\
&+&\rho_{T}(x,y)+\rho_{SDT}(x,y)+\rho_{IBT}(x,y),
\end{eqnarray}
where
\begin{eqnarray}
\rho_{IB}&=&IB(x,y),
\nonumber\\
\rho_{SD}&=&a^2\left[(1+\gamma)^2 SD^+(x,y)
+(1-\gamma)^2 SD^-(x,y)\right],
\nonumber\\
\rho_{IBSD}&=&2a\sqrt{r}\left[(1+\gamma) G^+(x,y)
+(1-\gamma) G^-(x,y)\right],
\nonumber\\
\rho_{T}&=&a^2~T(x,y),
\nonumber\\
\rho_{SDT}&=&2a^2\sqrt{r}\left[(1+\gamma) J^+(x,y)
+(1-\gamma) J^-(x,y)\right],
\nonumber\\
\rho_{IBT}&=&2a~I(x,y).
\end{eqnarray}
The explicit forms of the functions $IB(x,y)$, $SD^\pm(x,y)$,
$G^\pm(x,y)$, $T(x,y)$, $J^\pm(x,y)$ and $I(x,y)$ are given in
the Appendix.
The constant
\begin{equation}
a=\frac{m^2_\pi}{2f_\pi m_e}F_V=\frac{m^3_\pi}{8\pi^2f^2_\pi m_e}
\approx 3.945
\end{equation}
defines the strength of IB contribution relative to other contributions.

\section{Discussion and Conclusions}

\begin{figure}[th]
\mbox{\hspace{-0.3cm}}\epsfig{file=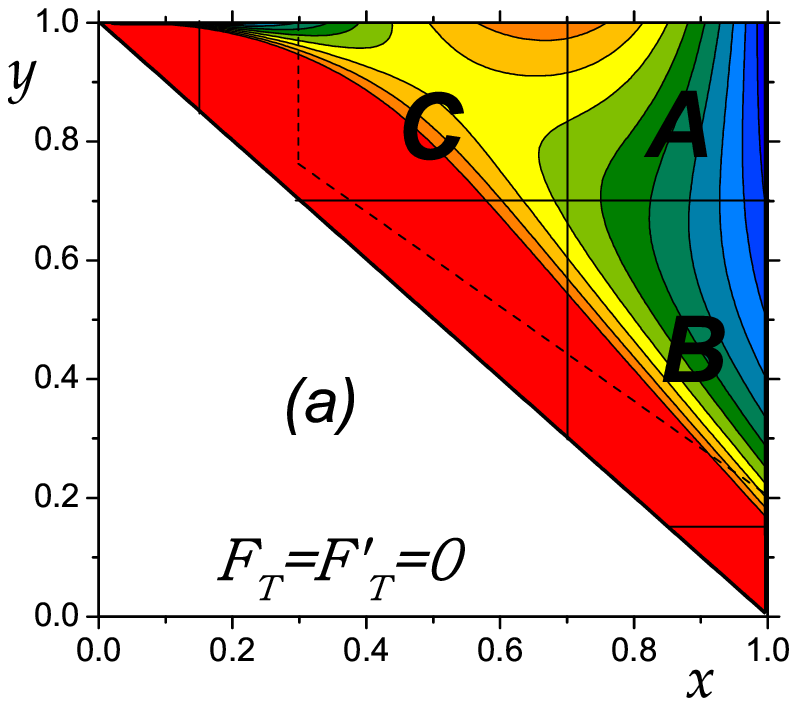,height=4.5cm,width=4.5cm}
\mbox{\hspace{-0.3cm}}\epsfig{file=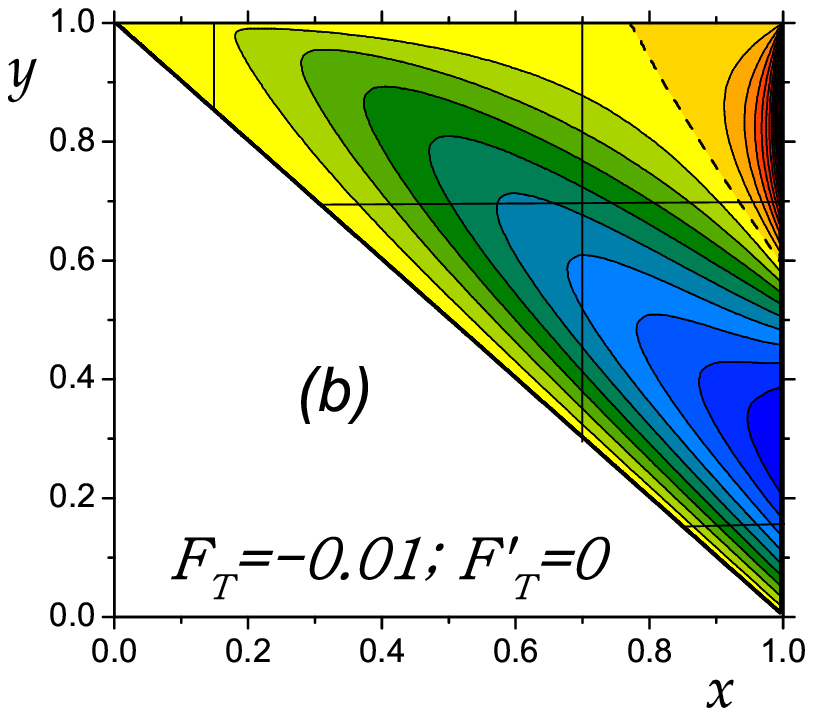,height=4.5cm,width=4.5cm}
\mbox{\hspace{-0.3cm}}\epsfig{file=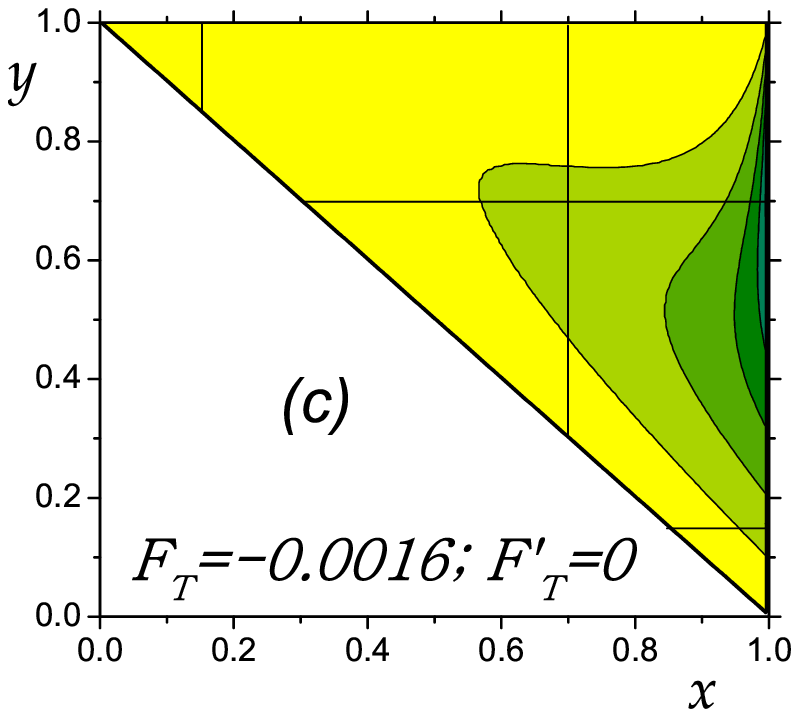,height=4.5cm,width=4.5cm}
\mbox{\hspace{-0.3cm}}\epsfig{file=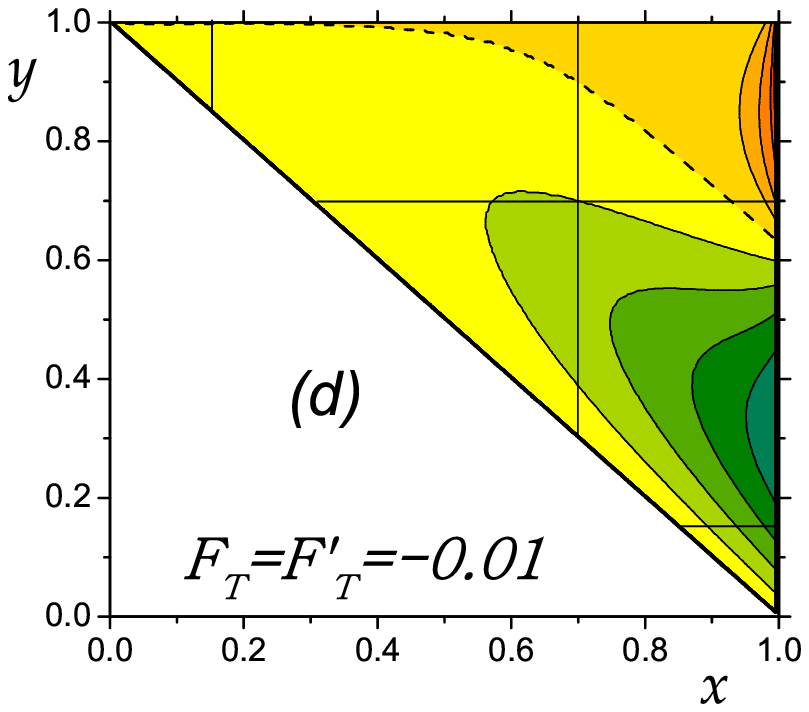,height=4.5cm,width=4.5cm}
\caption{(a): SM distribution (ISTRA region is marked out by dashed lines);
(b,c,d): isocurves (with 10\% step) corresponging to relative strength of tensor 
contribution for different values of $F_T$ and $F'_T$ (the dashed curve
corresponds to zero).}
\label{Dalitz}
\end{figure}

Based on the previous consideration, we discuss now the experimental data and 
their interpretation. To investigate the most interesting part of RPD, namely 
the SD radiation, and to extract $\gamma$, all previous experiments~\cite{old} 
have been fulfilled in a restricted kinematic region compatible with region $A$ 
of PIBETA experiment (Fig.~\ref{Dalitz}a), which is an intersection of regions $B$ 
and $C$.

In the Fig.~\ref{Dalitz}a the isocurves for pure SM contributions ($F_T$=$F'_T=0$) 
are shown. IB contribution dominates near the edge $x+y\simeq 1$ of the kinematically
allowed region and SD$^+$ reaches its maximum near the point $(2/3,1)$, 
while their interference and SD$^-$ contribution are small.

Almost the whole kinematically allowed region (Fig.~\ref{Dalitz}a) has been 
investigated first at ISTRA facility~\cite{ISTRA}. A large deficit of events 
$(33\pm 10)\%$ in comparison with SM prediction has been observed. Even with a poor 
statistics they were able to establish the kinematic region, where a lack of events 
occurs. This region corresponds to the bottom part of region $B$ of PIBETA experiment,
where only the IB and the small SD$^-$ contributions are expected.

The introduction of the tensor matrix element $M_T$ (\ref{MT1}) with
$F_T \simeq - 0.01$~\cite{Poblaguev,FT} explains the lack of events
and leads to a considerable improvement of the $\chi^2$. However, such a big 
tensor form factor contradicts~\cite{Voloshin} the present experimental data. 
Moreover, in order to explain the $(19.1\pm 2.5)\%$ deficit of events in region $B$ 
of PIBETA experiment, using the same form (\ref{MT1}) for the tensor matrix element,
an order of magnitude smaller value of the tensor form factor $F_T=-0.0016(3)$ 
is required~\cite{Dinko}. These different $F_T$ values indicate an inadequate 
description of the new interaction.

To analyse the problem further, we compare the relative strength of the tensor 
contribution with respect to SM contribution for two different values 
of the tensor form factor $F_T$ (Fig.~\ref{Dalitz}b and \ref{Dalitz}c).
The main difference between these two plots is the following: At the biggest value 
of $\vert F_T\vert$ both negative and positive contributions are present depending 
on the region, while for $\vert F_T\vert \lesssim 2f_\pi m_e/m^2_\pi\approx 0.0069$
the tensor matrix element leads to the only negative contribution into
the whole kinematically allowed region.

As one can see from the Fig.~\ref{Dalitz}b the tensor contribution with the biggest 
value $\vert F_T\vert$ affects significantly regions $A$ and $C$, besides region $B$. 
The latter is in contradiction with PIBETA results. The lowest absolute value of $F_T$
allows to evade this problem but leads to another latent difficulty of $\gamma$ 
determination. 

Indeed PIBETA claims two different values for the ratios of axial to vector form 
factors $\gamma=0.443\pm 0.015$ and $\gamma=0.480\pm 0.016$. The first one 
corresponds to the SM fits~\cite{PIBETA} to the entire data set, while the second -- 
to region $A$ data only. However, the $8\sigma$ deviation in the branching ratio 
from SM prediction in region $B$ requires an introduction of 
the new tensor contributions for proper $\gamma$ determination.
Then taking into account the correction for the destructive interference, 
the experimental branching ratio in region $A$ should be increased.
This increases the $\gamma$ value, as follows from the top panel of Fig.~4
in ref.~\cite{PIBETA}. However, if we believe in CHPT calculations~\cite{CHPT} 
$\gamma$ should be decreased in order to approach the lowest value. 
In other words the fits made with $F'_T=0$ are inappropriate.

The case $F_T=F'_T\simeq -0.01$ allows to describe both ISTRA and PIBETA anomalies 
as well, without contradiction to the experimental data. The corresponding tensor
contribution has a slight slope near its zero value (Fig.~\ref{Dalitz}d) in regions 
$A$ and $C$ which is in accordance with PIBETA results. Moreover, contrary to PIBETA 
fit, mainly the positive contribution in region $A$ leads to the right direction for 
$\gamma$ correction, corresponding to decreasing its value. 
While the tensor contribution can lead up to 40\% deficit in region $B$. 

The nonzero form factors $F_T$ and $F'_T$ indicate an existence of the quark-lepton
tensor interactions (\ref{R}) with the coupling constant $f_T=f'_T\simeq 0.013$.
Although such type of interactions can be generated through radiative corrections,
it is impossible to get the tensor coupling constant to be larger than
$10^{-9}-10^{-8}$ in SM and $10^{-4}-10^{-3}$ in its SUSY extensions~\cite{SUSY}.
Moreover, the particular form (\ref{R}) cannot arise as a result of 
Fierz transformations from a leptoquark exchange~\cite{Herczeg} as well.
The only natural source to produce the effective interaction (\ref{R}) is
the exchange of the new {\em chiral bosons}~\cite{MPL}, interacting 
{\em anomalously} with matter.

\section*{Acknowledgements}
I would like to thank A.~E.~Dorokhov who drew my attention to the preliminary
results of PIBETA experiment and D.~P.~Kirilova for the overall help.

\section*{Appendix}
The analytical expressions for the functions $IB(x,y)$, $SD^\pm(x,y)$,
$G^\pm(x,y)$, $T(x,y)$, $J^\pm(x,y)$ and $I(x,y)$ read
\begin{eqnarray}
IB(x,y)&=&\frac{y_1(1-r)}{x^2(x-y_1)}
\left[\frac{x^2}{1-r}+2(1-x)-\frac{2rx}{x-y_1}\right],
\nonumber\\
SD^+(x,y)&=&(x-y_1)\left[(x-y_1)(1-x)-rx\right],
\nonumber\\
SD^-(x,y)&=&y_1\left[y_1(1-x)+rx\right],
\nonumber
\end{eqnarray}
\begin{eqnarray}
G^+(x,y)&=&\frac{y_1}{x(x-y_1)}
\left[(x-y_1)(1-x)-rx\right],
\nonumber\\
G^-(x,y)&=&\frac{y_1}{x(x-y_1)}
\left[y_1(1-x)-(1-r)x\right],
\nonumber\\
T(x,y)&=&2\left[(\gamma_T-\gamma'_T)^2+\gamma^2_T\right]y_1(x-y_1),
\nonumber\\
&&+~\gamma'^2_T\frac{rx}{1-x}\left[x-2y_1-\frac{rx}{1-x}\right],
\nonumber\\
J^+(x,y)&=&-\gamma'_T x\left[x-y_1-\frac{rx}{1-x}\right],
\nonumber\\
J^-(x,y)&=&(\gamma'_T-2\gamma_T)xy_1,
\nonumber\\
I(x,y)&=&\gamma'_T y_1+2(\gamma_T-\gamma'_T)y_1
\left[\frac{1}{x}-\frac{r}{x-y_1}\right]\!,
\end{eqnarray}
where $y_1=1-y+r$, $\gamma_T=F_T/F_V$ and $\gamma'_T=F'_T/F_V$.


\end{document}